\documentclass[review]{elsarticle}

\usepackage{lineno,hyperref}
\modulolinenumbers[1]

\journal{arXiv.org}
\usepackage[utf8]{inputenc}
\usepackage{amsmath,amsthm,amsfonts}
\usepackage{enumitem}
\usepackage{graphicx}
\usepackage{color}
\usepackage{float}
\usepackage{fixltx2e} 
\usepackage{placeins}
\graphicspath{ {./}}
\usepackage{geometry}
\newcounter{fig_num} 
\begin{document}
\begin{frontmatter}
\title{Automated segmentation and morphological characterization of placental histology images based on a single labeled image}

\author[address1]{Arash Rabbani \corref{mycorrespondingauthor}}
\cortext[mycorrespondingauthor] {Corresponding author}
\ead{a.rabbani@leeds.ac.uk | rabarash@gmail.com}

\author[address2]{Masoud Babaei}
\author[address3]{Masoumeh Gharib}

\address[address1]{The University of Leeds, School of Computing, Leeds, UK}
\address[address2]{The University of Manchester, School of Chemical Engineering and Analytical Science, Manchester, UK}
\address[address3]{Mashhad University of Medical Sciences, Department of Pathology, Mashhad, Iran}

\begin{abstract}

In this study, a novel method of data augmentation has been presented for the segmentation of placental histological images when the labeled data are scarce. This method generates new realizations of the placenta intervillous morphology while maintaining the general textures and orientations. As a result, a diversified artificial dataset of images is generated that can be used for training deep learning segmentation models. We have observed that on average the presented method of data augmentation led to a 42\% decrease in the binary cross-entropy loss of the validation dataset compared to the common approach in the literature. Additionally, the morphology of the intervillous space is studied under the effect of the proposed image reconstruction technique, and the diversity of the artificially generated population is quantified. Due to the high resemblance of the generated images to the real ones, the applications of the proposed method may not be limited to placental histological images, and it is recommended that other types of tissues be investigated in future studies. The image reconstruction program as well as image segmentation model are available publicly.

\end{abstract}

\end{frontmatter}


\section{Introduction}
\subsection{Human placenta}
\begin{figure}[H]
	\centering\includegraphics[page=\value{fig_num},width=.8\linewidth]{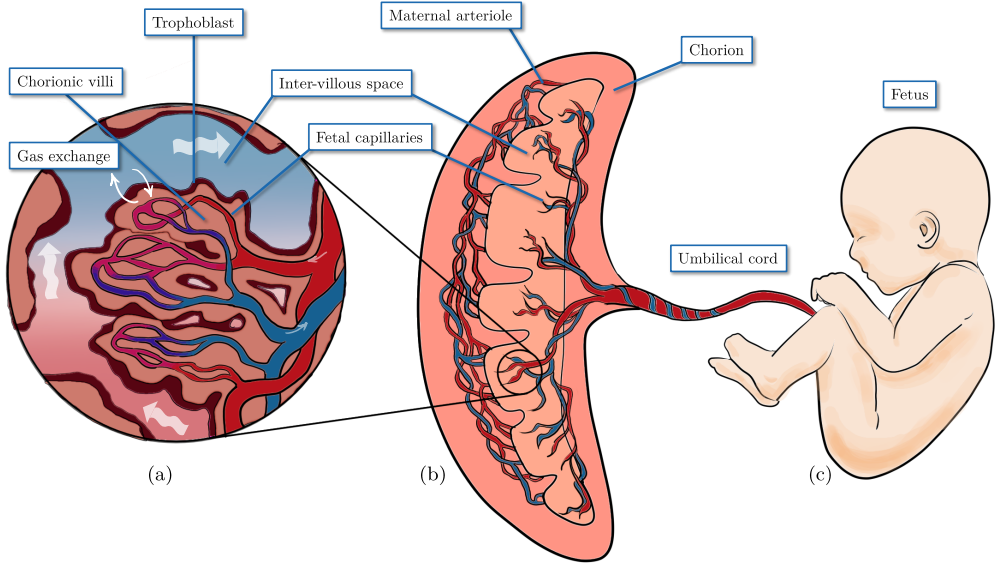} \stepcounter{fig_num}
	\caption{Placental internal structure and its role in delivering oxygen and nutrients to the fetus. a) internal structure of chorionic villus including fetal capillaries and trophoblast which acts as a selective membrane between maternal and fetal blood, b) cross-section of placenta structure including umbilical cord, maternal arteriole, intervillous space, and chorion which is the outermost membrane around the fetus, c) fetus.}
	\label{fig:placenta_intro}
\end{figure}

During pregnancy, the placenta is a temporary but vital organ in female humans that connects the baby to the uterus \cite{burton2015placenta}. The placenta forms soon after conception and adheres to the uterine wall on one side and connects to the baby on the other side via the umbilical cord \cite{burton2015placenta} as visualized in Fig. \ref{fig:placenta_intro}-c. The role of the placenta includes providing a passage for the transport of oxygen, nutrients, hormones, and immune cells. The umbilical cord bifurcates into numerous vessels and capillaries that reside in the villous space of the placenta (Fig. \ref{fig:placenta_intro} -b). Fetal capillaries pass close to the maternal blood pool in the intervillous space, and controlled transport of oxygen and nutrients occurs at the cellular level through a set of selective membranes known as the syncytiotrophoblast and cytotrophoblast which collectively are called the trophoblast\cite{cole2020100,wang2010vascular} (Fig. \ref{fig:placenta_intro}-a). In several studies, the microstructural morphology of the villous space has been found to have a significant correlation with the healthiness and development of the fetus \cite{balihallimath2013placental, winship2015blocking, nelson2009placental}. Such observations justify the need to take a closer quantitative look at the morphology of the placenta for diagnostic and prognostic purposes.

\subsection{Placental histology}
Microscopical study of histological slides provides an invaluable visual representation of tissues and cells on the micrometer scale. Histological images play a critical role in the study of body functions and the diagnosis of certain diseases \cite{hussein2015once} from acute and chronic inflammation \cite{li2018inflammatory,shah2017histopathologic} to autoimmunity \cite{deheragoda2007use} and cancer \cite{roux2013mitosis}. In the same way, these images have been frequently used to shed light on the micro-structural morphology of the placenta \cite{wislocki1943histology}.  Several maternal conditions and diseases can affect the structure and functionality of the placenta, which can eventually lead to adverse effects on the fetus development \cite{facchetti2020sars,nelson2009placental}. As an example, using histo-pathological images in combination with other imaging techniques, Facchetti \textit{et al. } \cite{facchetti2020sars} observed that SARS-CoV2 infection in mothers can cause neutrophil infiltration in the intervillous spaces and consequently adversely affect the performance of the placenta. As another example, type 1 diabetes has been found to cause abnormal placentas that are enlarged, thick, and plethoric, with abnormalities of villous maturation \cite{benirschke2012maternal,nelson2009placental}. 

Placental histological samples are commonly obtained by biopsy in the chorionic villi section. Sampling is usually performed in the 11th to 14th weeks of pregnancy \cite{ong2004maternal} using a thick needle or a small tube through the abdomen or cervix, respectively. The recovered samples are fixed with paraffin or resin and some slices are cut from the structure with a thickness of a few microns that helps light pass through \cite{macenko2009method}. Using light microscopy, these thin slices can be visualized for qualitative and quantitative studies on placenta health and function. To perform quantitative studies on placental histological images, manual or automated data analysis methods are used. Considering the fact that the manual analysis of placental histological slides by microscope is a costly and time-consuming activity, computer-aided analyzes offer a fast and repeatable result that eliminates inter- and intra-observer variability \cite{salsabili2019automated}. 

\subsection{Image synthesis}
Medical images are expensive to obtain and process both in terms of budget and time. A quick solution to proliferate a variety of desired images at virtually no cost is image synthesis. Synthetic images have been widely approached in litrature to help for automated diagnosis and classification of several diseases from liver lesion \cite{frid2018gan} and cardiac abnormalities \cite{prakosa2013cardiac} to brain tumors\cite{shin2018medical}. Such an approach relies on a set of exemplar images to be used for training and then by capturing the main elements within the images a diversified set of predicted images are generated that have a certain feature in common \cite{xue2021selective,wei2019generative,portenier2020gramgan,tahmasebi2016enhancing}. Patch-based reconstruction of textural images as one of the classic image synthesis approaches have been used in the present study to augment the availible image dataset \cite{efros1999texture,efros2001image}. We have tailored the method to fit the purpose of this study to regenerate histological images of chorionic villi. 

\subsection{Application of deep learning}
Image segmentation is the process of dividing an image into its structural elements that share common characteristics \cite{pham2000current}. Segmenting placental histological images into villous and intervillous spaces is a fundamental task in studying the morphology of the placenta \cite{swiderska2018image}. Automated methods have been used for this purpose to save time and eliminate operator bias from the diagnostic cycle \cite{salsabili2019automated,mobadersany2021gestaltnet}. 

In addition to intervillous space segmentation, deep learning has been used for cellular phenotyping based on placental histology images by Ferlaino \textit{et al.} \cite{ferlaino2018towards}. For this purpose, they have created a dataset composed of thousands of high-confidence, manually curated placenta cells from five classes. By combining a nuclei-locator deep learning model and an image-based classifier, they have been able to achieve an accuracy of around 75\% for the detection of different types of placental cells. Deep learning can be also used for improving the quality of the placental histology images as shown by Rabbani and Babaei \cite{rabbani2022resolution}. They have developed a convolutional neural network model that takes low-resolution as input and predicts the a residual image that shows the differences between the low- and desired high-resolution images. They have shown that such an approach does not only intensifies main feautres of the image but alos, adds realistic-look details that can helps for better understanding of the low quality images of placenta.      

Another recent application of deep learning on placental histological images has been presented by Mobadersany \textit{et al.}\cite{mobadersany2021gestaltnet}. Their developed deep learning model known as \textit{GestaltNet} can estimate gestational age by analyzing the scanned placental slides. They have been able to estimate gestational age by a mean absolute error of 1.0847 weeks using a deep learning method known as attention-based feature aggregation \cite{mobadersany2021gestaltnet,kim2018regional}. To generate the required dataset for training, validation, and testing the \textit{GestaltNet}, the authors manually annotated 1918 regions on 154 histological slides \cite{mobadersany2021gestaltnet}, which is a considerably time-consuming task. 

Common deep learning techniques rely on a relatively large dataset of manually segmented images for training purposes. In the present study, we have minimized the requirement of manually segmented images by relying on a single image for the training process. This task is done through an innovative data augmentation technique that creates a diversified range of tissue textures that enables the model to perform at an acceptable level of accuracy in the prediction of unforeseen image datasets with significantly different textures.

\section{Methodology}
\label{sec:me}
This study presents a tailored method of histological data augmentation for AI--powered segmentation of the placental intervillous space. We expect that the presented method is especially effective in the scarcity of labeled data. Labeling histological images is an expensive and time--consuming task, especially when the presence of highly trained specialists is crucial. In this section, we initially describe the available data, and then introduce the proposed image augmentation technique. In addition, the structure of four deep learning models used for image segmentation is discussed. Finally, an image--based quantitative method is presented that helps in characterizing the morphology of the segmented placental intervillous space. 
\subsection{Material}
\begin{figure}[H]
	\centering\includegraphics[page=\value{fig_num},width=.8\linewidth]{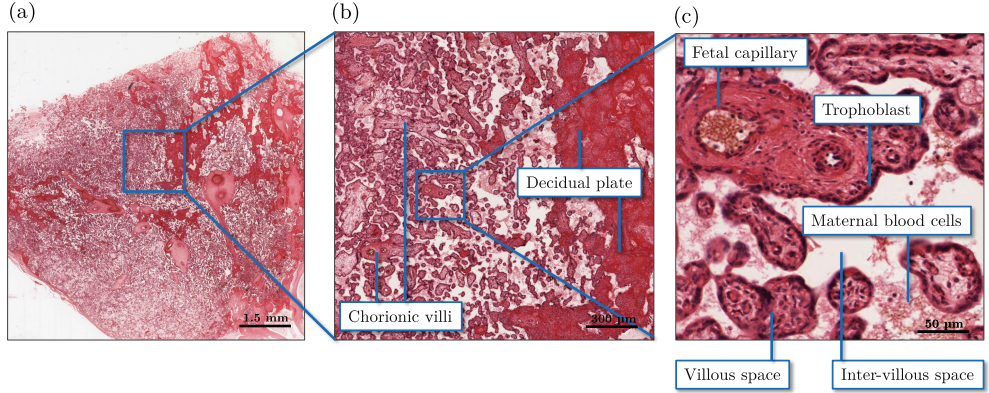} \stepcounter{fig_num}
	\caption{Histological image of the placenta from a healthy volunteer in three length scales, a) overview of the sample which shows maternal to fetal tissues extended from right to left, b) chorionic villus inter-twined with intervillous space in the vicinity of decidual plates, c) 40X magnified view of chorionic villus including fetal capillaries and syncytiotrophoblast, as well as intervillous space with the presence of maternal blood cells.}
	\label{fig:placenta_histology}
\end{figure}

In this research, H \& E histological images of three placenta samples from healthy volunteers (HV) have been used to demonstrate a possible improvement in automated segmentation techniques. The images were obtained from the University of Michigan Histology and Virtual Microscopy archive \cite{hortsch}. In total, 34 zoomed images with a size of $256 \times 256$ pixels have been acquired from the original 3 histological slides with a maximum magnification of $40X$. All zoomed images were captured from locations with the presence of chorionic villus and intervillous space. Fig. \ref{fig:placenta_histology} illustrates three magnifications of the histological image obtained from HV $\#1$. Fig. \ref{fig:placenta_histology}--a shows the complete slide with maternal to fetal tissues extended from right to left. By zooming five times, Fig. \ref{fig:placenta_histology}--b  is obtained, which illustrates the chorionic villus intertwined with intervillous space near the decidual plates, which are compact maternal tissues protecting the fetus which should not be included in the chorionic villus in the segmentation process. Finally, Fig. \ref{fig:placenta_histology} --c shows the 40X magnified view of the chorionic villus that includes the fetal capillaries and syncytiotrophoblast, as well as intervillous space with the presence of maternal blood cells that should be included as intervillous space in the segmentation process.  

\subsection{Data augmentation}
\begin{figure}[H]
	\centering\includegraphics[page=\value{fig_num},width=.8\linewidth]{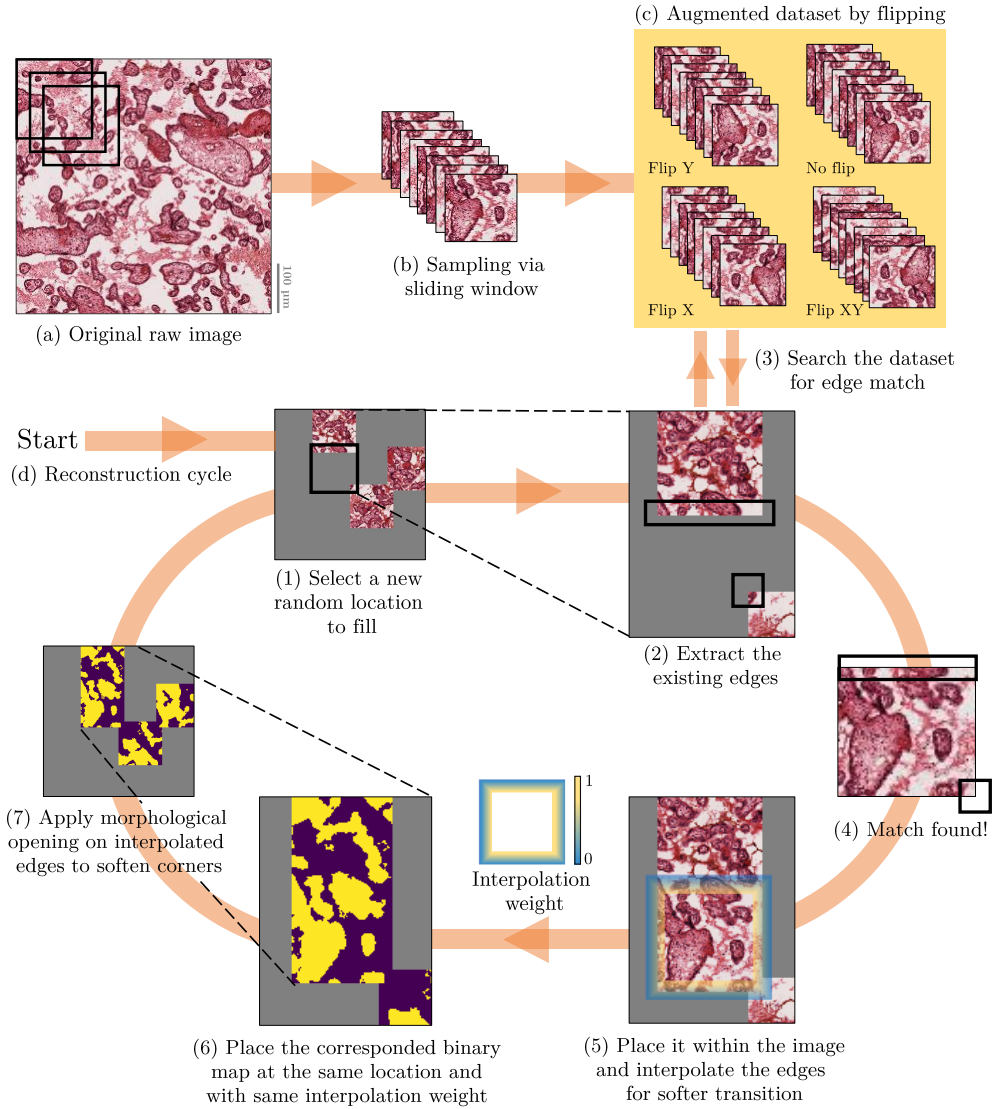} \stepcounter{fig_num}
	\caption{Steps of the proposed image reconstruction process from making a dataset of sub-samples to reconstruct a new realization. a) the original input image, b) splitting the input image into overlapped sub-samples, c) flipping the obtained dataset for augmentation purpose, d) the cycle of constructing a new realization by searching through the dataset for matching edges and fusing the obtained match for a softer transition.}
	\label{fig:workflow}
\end{figure}

\begin{figure}[H]
	\centering\includegraphics[page=\value{fig_num},width=.8\linewidth]{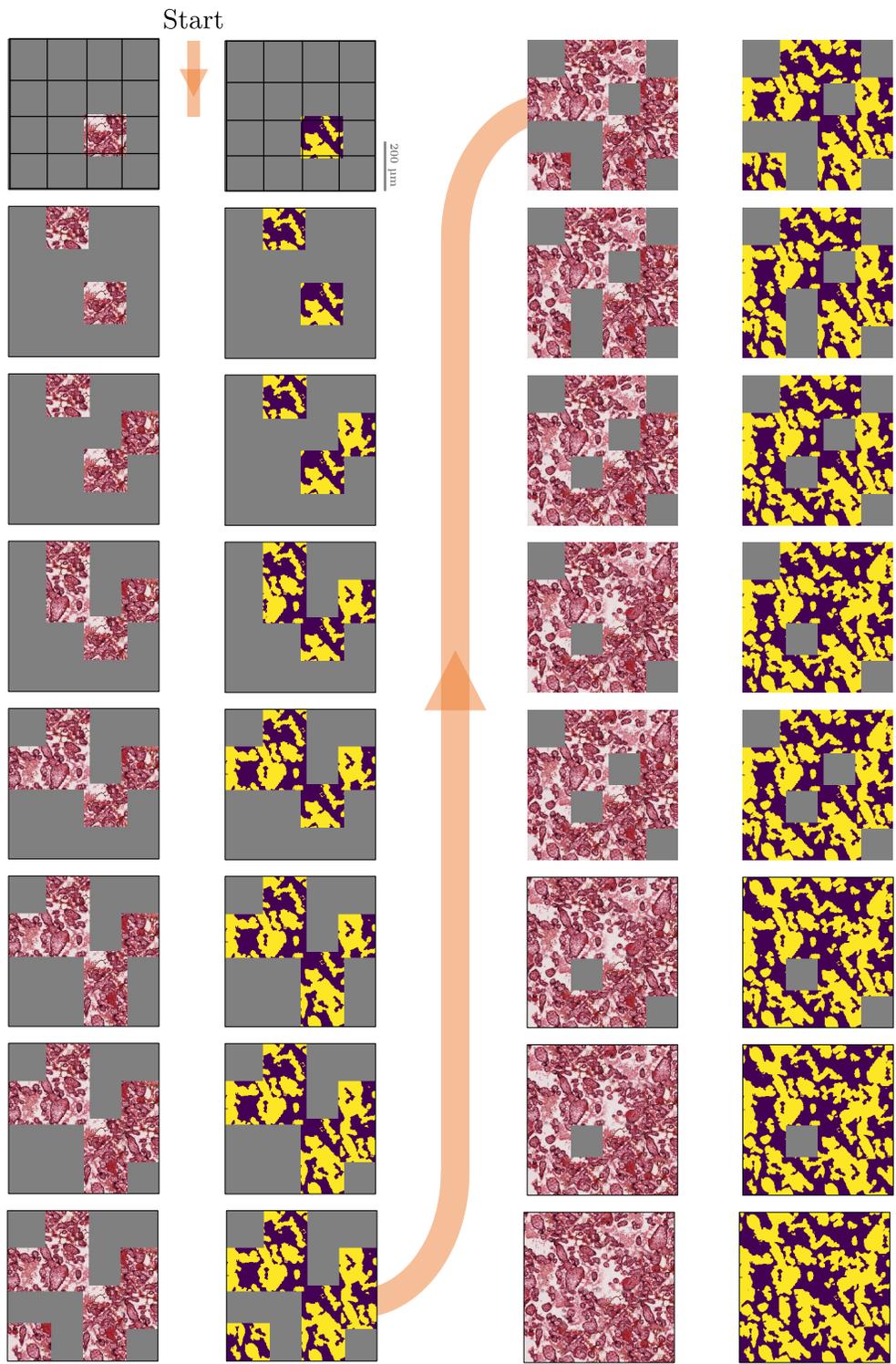} \stepcounter{fig_num}
	\caption{An example of a reconstruction using a $4 \times 4$ grid starting from a random tile selection at the first step to generating a complete realization of placental histology image along with the segmented maps of villous space that get complete based on the corresponding tile from the histology image.}
	\label{fig:an_example}
\end{figure}

Considering the aim of this study to offer automated segmentation of placenta slides with minimal availability of labeled data, we have selected only one of the $256 \times 256$ histology images of HV $\#1$ to act as our training image (Fig. \ref{fig:workflow}--a). The rest of the 33 available images will be used for validation purposes. Considering the high demand of deep learning models for training data, artificial augmentation of the available datasets is an inevitable step \cite{shorten2019survey}. The current state-of-the-art in augmentation of histological images includes color shift, noise addition, zooming, rotation, flipping, translating, and elastic deformation \cite{tellez2019quantifying,faryna2021tailoring} which is called \textit{base case} in this paper. In this section, our aim is to develop an additional step in relation to previous methods of image data augmentation to improve the diversity of the generated data, which can lead to higher model performance. Fig. \ref{fig:workflow} describes the steps required in the proposed data augmentation method. Initially, a dataset of sub--sample images is cropped from the original image to a smaller size via a sliding-window approach. The sliding sampling window sweeps the area of the original image with discrete 5 pixel steps in each direction (Fig. \ref{fig:workflow}-a). Then, the obtained dataset of sub-samples (Fig. \ref{fig:workflow}-b) is augmented by flipping in horizontal (X), vertical (Y), and a combination of both directions (XY) (Fig. \ref{fig:workflow}-c). Now, the reconstruction cycle begins by selecting a new random location in the new image to fill (Fig. \ref{fig:workflow}--d--1). Each of the locations is considered to have 5 pixels overlap with neighboring locations similar to the sampling stage. If the edges of the selected location have not been occupied by neighbor blocks in advance, we are free to select any of the sampled images in the dataset to fill that specific location. In contrast, if the selected location has predetermined edges from previous cycles, we need to look up in the created dataset of sub--samples (Fig. \ref{fig:workflow}-c) for entries with matching edges (Fig. \ref{fig:workflow}--d--3). The search is done by temporarily masking all parts of the image except the edges and then minimizing the mean squared error for the pixel values of the edges of the image compared to the dataset. To maintain the texture and alignment of the original image, the search within the \textit{no flip} part of the dataset has a higher priority (Fig. \ref{fig:workflow}-c). Matches are not perfect, and most of the time inconsistencies still exist at the interface of two adjacent blocks. In this regard, we have proposed a weighted cubic interpolation between the edge pixels of the matched image and pre--existing pixels on the image (Fig. \ref{fig:workflow}--d--5). The interpolation layer has a width of 10 pixels starting from the inside of the matched sub-sample to 5 pixels beyond its borders to offer a soft transition. As a follow-up to histology image reconstruction, the same process is replicated for the segmentation mask that separates villous and intervillous spaces using ones and zeros, respectively (Fig. \ref{fig:workflow}--d--6). To soften the corners of the mask image, morphological image opening is recommended for the pixels that are located beneath the non--zero elements of the interpolation layer. Finally, by repeating the cycle, all parts of the image grid will be completed, and a new realization of the original image is created. Fig. \ref{fig:an_example} illustrates the output of the cyclic image filling for an image using a $4 \times 4$ grid. It is noteworthy that the search process becomes more computationally intensive when approaching the completion of the image with more pre-existing edges that need to be matched and satisfied.                 

\subsection{Image segmentation via deep learning}
In this section, the objective is to train a deep learning model that distinguishes between villous and intervillous space in placental histology images. For this purpose, we have selected four deep learning networks that have been frequently used for similar segmentation purposes in the literature, comprising U-net \cite{ronneberger2015u}, Res-net \cite{wu2019wider,lu2021wbc}, ResU-net \cite{li2019residual,kalapahar2020gleason}, and a simple fully convolutional model (FC-net) \cite{tran2016fully} without skip or residual connections. The utilized U-net structure is composed of 41 layers with around $1350 K$ trainable parameters, activation functions of the Exponential Linear Unit (ELU), and \textit{He Normal} \cite{he2015delving} kernel initializer. The convolutions have had a filter size of $3 \times 3$ with the same-size padding and stride of $2 \times 2$. More information on the structure and performance of the U-net can be found in \cite{ronneberger2015u}. 

Res-net model in this study has 82 layers with around $833 K$ trainable parameters. There are three residual connections in the encoding section of the model, followed by one at the bottleneck and three in the decoding section. Two-dimensional convolutions have a filter size of $3 \times 3$ with the same-size padding and Rectified Linear Activation Units (ReLU). 

The utilized ResU-net model is fundamentally similar to the U-net model with the difference of having 8 residual units divided equally between the encoding and decoding sections of the model. As a result, this model will have 57 layers and around $1750K$ trainable parameters. The rest of the features of the model are identical to the discussed U-net structure. 

In addition to the discussed complex models, we have included a more simple convolutional neural network model as well to verify the improvements due to the proposed data augmentation technique. A fully convolutional model with 41 total layers and around $1380 K$ trainable parameters is selected. This network has a filter size of $3 \times 3$ with the same-size padding and ReLU activation function.

We have trained all the deep learning models mentioned from scratch using the binary cross entropy loss function and Adam optimizer with a learning rate of 0.0005, with the exponential decay rate for the first moment equal to 0.9 and the exponential decay rate for the second moment equal to 0.999. 

After developing the structure of the deep learning models, we have used the two discussed augmented datasets to start training and compare the results. The input of the models are colored H \& E histological images with the size of $256 \time 256$ pixels and a depth of three to represent the red, green and blue channels. Furthermore, the output of the models will be an array of $256 \time 256$ pixels with values between 0 and 1 representing the likelihood that a pixel will be in the villous space. A thresholding filter will be applied to all model outputs to discretize the values into 0 and 1, to indicate intervillous and villous spaces, respectively.   

\subsection{Morphological characterization}
\begin{figure}[H]
	\centering\includegraphics[page=\value{fig_num},width=.8\linewidth]{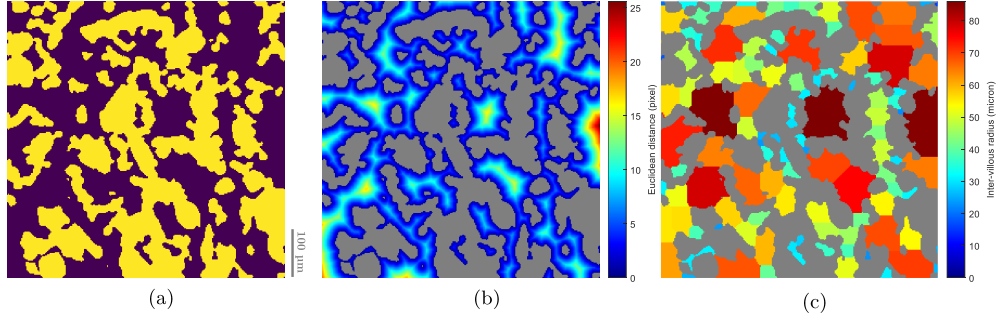} \stepcounter{fig_num}
	\caption{An example of intervillous space chambers segmentation using watershed algorithm, a) Binary input image, b) Distance map of the intervillous space, and c) segmented instances of intervillous chambers colored based on the radius of the maximal inscribed circles.}
	\label{fig:watershed}
\end{figure}

In order to provide a more in-depth comparison between the results of the training of the model using the base case and the proposed augmentation methods, the morphological properties of the segmented placenta images are investigated. The morphological properties of the placental geometry can give a good measure of fluid conductance and mass transfer behavior, which can directly affect the health of the fetus \cite{nelson2009placental}. One of the important properties we aim to calculate is the specific surface of the intervillous space, which is equivalent to the ratio between the internal surface area and the total volume of the intervillous space \cite{nelson2009placental}. In addition, the size distribution of the intervillous pathways and the average connectivity of the intervillous network are investigated. The average connectivity of a network is defined as the mean number of links connected to each of the nodes in the network, and in these terms we can assume an inter-connected network structure for the intervillous space that carries the maternal blood. The method we have used to characterize the intervillous network structure is known as the watershed segmentation algorithm. The watershed segmentation algorithm is mainly known for separating attached objects based on their morphology \cite{sheppard2004techniques,Rabbani2014}, and in this study, our aim is to divide the intervillous space into smaller chambers and measure the connectivity between chambers to have a quantitative measure of the morphology of the intervillous space. Watershed segmentation algorithm has been widely used to build simplified pore network models when dealing with heterogeneous porous material \cite{rabbani2021image}. In a sense, if we assume the fetal section of the placenta to act as a solid space and intervillous parts as a void space, a porous structure can be considered for placenta, and many of the previously developed methods in the field of porous material research can be tailored and used to characterize and model the internal micro-structure of placenta. 
In the watershed segmentation algorithm, initially the distance map of the binary geometry (Fig. \ref{fig:watershed}--a) will be calculated, which shows the distance between each of the pixels in the void space and the closest solid wall (Fig. \ref{fig:watershed}--b). In analogy, consider that the generated distance map mimics a geological landscape with distance values that represent the depth of the land. Now, if rain falls on such a landscape, water will accumulate at the deepest points and form small lakes. If we keep track of the different lakes while the water level is rising, they would represent separated intervillous chambers (Fig. \ref{fig:watershed}--c). This process continues until the whole distance map is flooded with water, and that is where the term "watershed" comes from. In the next step, we can run a neighborhood analysis to find the adjacent chambers and build a network based on the chamber inter-connectivity. Such a network can be used for characterization and modeling purposes and help to mathematically describe the fluid flow behavior within the intervillous space. More details on the network extraction process can be found in \cite{rabbani2019hybrid,Rabbani2014,baychev2019reliability}.

\section{Results and Discussions}
In this section, we have initially discussed the results of data augmentation using the proposed method in terms of intervillous morphological properties. Then, the performance of the discussed deep learning models is studied to evaluate the effectiveness of the developed data augmentation technique. In addition, the morphology of the predicted intervillous maps are compared with the ground-truth binary image. Furthermore, we have studied the effect of changing the size of the grid in the process of image reconstruction. Finally, some discussions are presented regarding future applications of the presented data augmentation technique. 

\subsection{Data augmentation}
\begin{figure}[H]
	\centering\includegraphics[page=\value{fig_num},width=.8\linewidth]{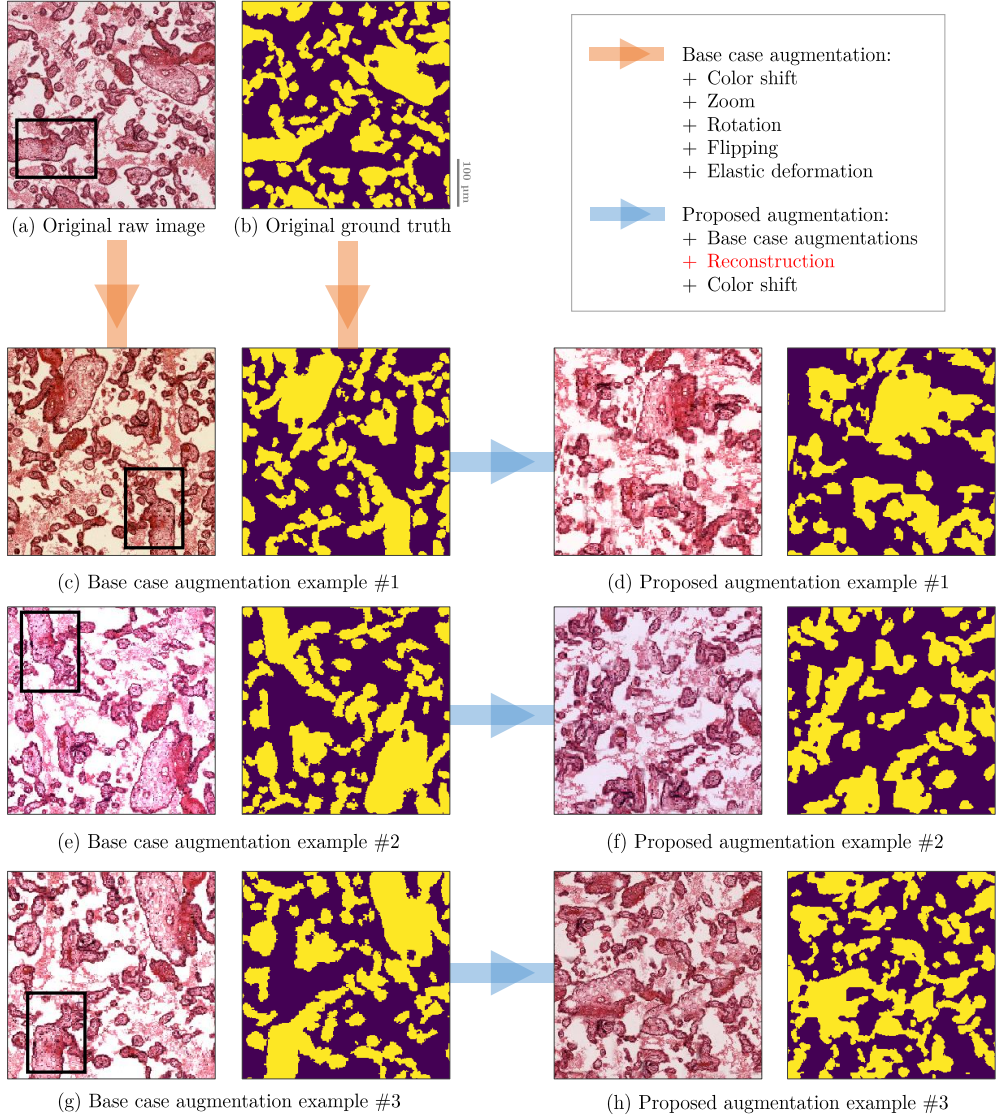} \stepcounter{fig_num}
	\caption{Comparison between the base case augmentation composed of color shift, zooming, rotation, flipping and elastic deformation which are the current state-of-the-art method for augmenting the medical images (c, e, and g), and the proposed method which is composed of the base case method followed by the image reconstruction technique presented in this paper (d, f, and h). The highlighted boxes show a specific chorionic villi which is distinguishable in all base case augmentations, which disappears in the proposed augmentations due to a more significant change of texture.}
	\label{fig:comparing_examples}
\end{figure}

To compare the performance of the training using the base case and the proposed data augmentation method, we have generated two image datasets using each of the methods with 3000 samples. All of these augmented samples are generated using a single raw image and its ground-truth label (Fig. \ref{fig:comparing_examples}-a and b). In the base case augmentation, we have used a randomized set of operations including color shift up to $\pm 10 \%$ for each channel, zooming up to 150\%, clockwise rotation of less than 10 degrees, horizontal and vertical flipping as well as a combination of them, and elastic deformation using the following equations:

\begin{equation}
	\label{eq:elasticx}
x_{new}=x+(0.025 \times S_1\sigma)\sin(x/(S_1/M) \pi+\pi R)
\end{equation}

\begin{equation}
	\label{eq:elasticy}
	y_{new}=y+(0.025 \times S_2\sigma)\cos(x/(S_2/M) \pi+\pi R)
\end{equation}

Where $S_1$ and $S_2$ are dimensions of the image in pixels, $\sigma$ is the standard deviation of the amplitude in elastic waves which we have considered to be a random number between 0 and 1, $M$ is the mesh size ratio and takes a random value between 1 and 5, and $R$ is a random number between 0 and 1 that causes a random phase change in the elastic wave by shifting the angle between 0 and $\pi$. Furthermore, $x$ and $y$ are the coordinates of the pixels in the image grid that will be relocated in $x_{new}$ and $y_{new}$, respectively. The values of the pixels for the deformed image are calculated using interpolation with the order of one for raw images and zero for the labeled maps. The result is an image with wave-like distortion in the vertical and horizontal directions, which helps to improve the diversity of the augmented dataset. For more information on elastic deformation, refer to \cite{castro2018elastic}.

The proposed method of reconstruction--based data augmentation can be assumed as a further step after the implementation of the base case augmentation techniques. Fig. \ref{fig:comparing_examples} illustrates 3 examples of the augmented dataset processed by the two methods discussed. The first column of the examples (Fig. \ref{fig:comparing_examples}-c, e, and g) is generated using the base case approach, while the second column images (Fig. \ref{fig:comparing_examples}-d, f, and h) have benefited from the additional step of image reconstruction. If we follow the highlighted black rectangle from the original image to the base case examples, it is not surprising that a specific villi structure has been repeated in all of the images, but with different rotations. This repetition will limit the ability of the AI model to learn new structures. On the contrary, the images that are further modified using the proposed reconstruction technique do not exhibit such a repetitive villi pattern, which theoretically can give the AI model the opportunity to learn from a wider range of structures.

\begin{figure}[H]
	\centering\includegraphics[page=\value{fig_num},width=.8\linewidth]{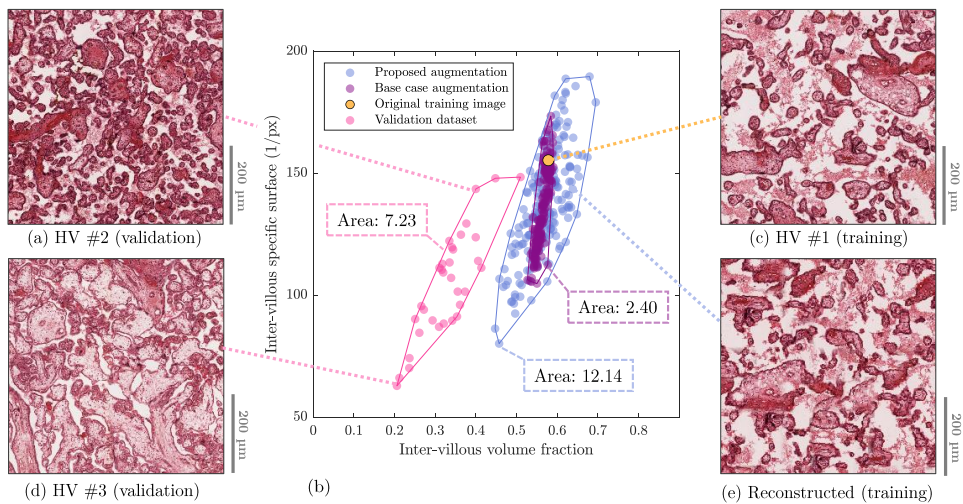} \stepcounter{fig_num}
	\caption{Comparing the augmentation variations in the domain of intervillous specific surface versus volume fraction. In addition, the training image and validation dataset are visualized. a) a validation image with high specific surface, b) 2--D representation of the datasets, c) training image, d) a validation image with low specific surface, e) an example reconstructed image using the proposed method.}
	\label{fig:domain}
\end{figure}

To visualize and compare the diversity of the generated data using both augmentation methods, we have implemented a dimension reduction technique by calculating two morphological features of the intervillous space. intervillous volume fraction and the intervillous specific surface have been calculated for 200 random samples from each data set and plotted against each other (Fig. \ref{fig:domain}-b). In addition, we have included the original training image as a single point, as well as the validation dataset as a group of data points in the graph. As can be seen in Fig. \ref{fig:domain}-b, the original training image is located outside the region occupied by the validation dataset, which indicates the dissimilarity of the training image and the validation dataset. Such a condition can lead to challenging model training, since the model is trained on an image which is not closely similar to the prediction set. However, in real-world examples, such issues are likely to occur since it is not feasible for a training dataset to include any possible diversity that exists in an out-of-the-sample prediction dataset. If comparing the area occupied by the proposed augmentation technique and the base case in Fig. \ref{fig:domain}-b, it can be concluded that the presented method is capable of adding diversity to the training dataset to the extent that it approaches the validation dataset which can improve the chance of correct predictions. 
To give a more physical sense in Fig. \ref{fig:domain}, some example images are provided (Fig. \ref{fig:domain}-a, c, d, and e). As can be seen, images with a lower intervillous volume fraction, such as Fig. \ref{fig:domain}-d, have a more compact arrangement of villous elements. In addition, images with larger intervillous specific surfaces have a higher number of isolated villous elements, which increases the surface area of the intervillous compared to its volume.

\subsection{Prediction performance comparison}
\begin{figure}[H]
	\centering\includegraphics[page=\value{fig_num},width=.8\linewidth]{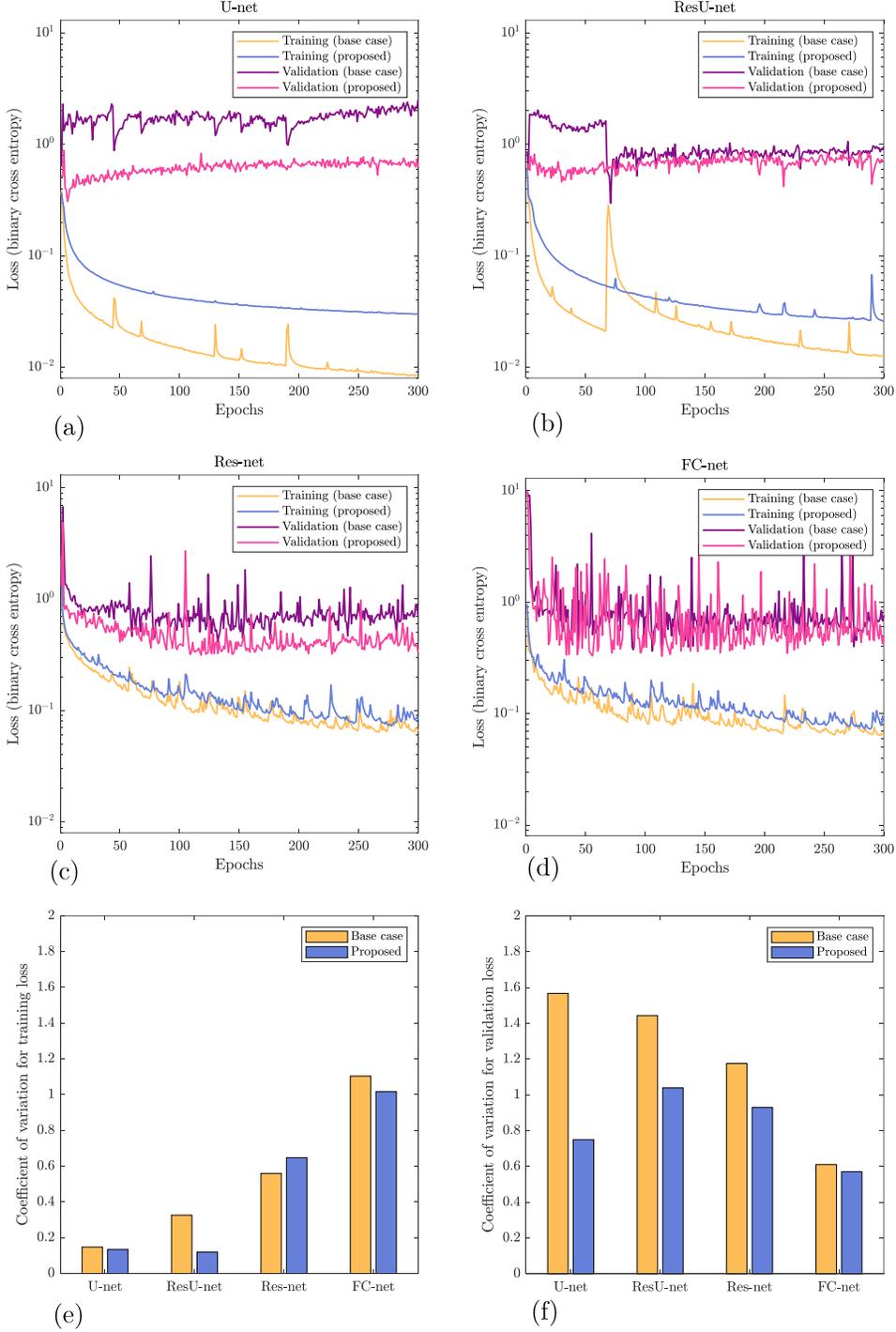} \stepcounter{fig_num}
	\caption{Comparing the training and validation binary cross--entropy losses for 4 types of the deep learning structures studied using base case and proposed methods of data augmentation (a to d), as well as the coefficient of variation for training (e) and validation (f) losses across different epochs. Significantly lower validation losses for all 4 tested network structures indicates a lower chance of over-fitting using the proposed augmentation approach.}
	\label{fig:performance_comp1}
\end{figure}

The two augmented image datasets have been used to train four different deep learning models with the purpose of semantic segmentation. The results of the training and validation performance for U--net, ResU-net, Res-net, and FC-net are visualized in Figs. \ref{fig:performance_comp1}-a to d. For all deep learning models, the proposed augmentation method has been observed to lead to a lower validation loss and a higher training loss. The lower training loss of the base case method is predictable considering the noticeable similarities and low diversity of the images in its dataset. On the other hand, a lower validation loss for the proposed approach is an indicator of improved performance for all AI models tested. 
In addition, it is observed that the training and validation losses of the proposed approach are declining with a more stable trend compared to the base case. This observation can be quantified by looking at the coefficient of variation for all the loss curves obtained, as visualized in Fig. \ref{fig:performance_comp1}-e and f. For all of the deep learning models, except for the training loss of the Res-net model, the coefficient of variation has been observed to be lower when using the proposed augmentation approach. A more stable loss curve can be an indicator of a well--optimized model and repeatability of the training process.

\begin{figure}[H]
	\centering\includegraphics[page=\value{fig_num},width=.8\linewidth]{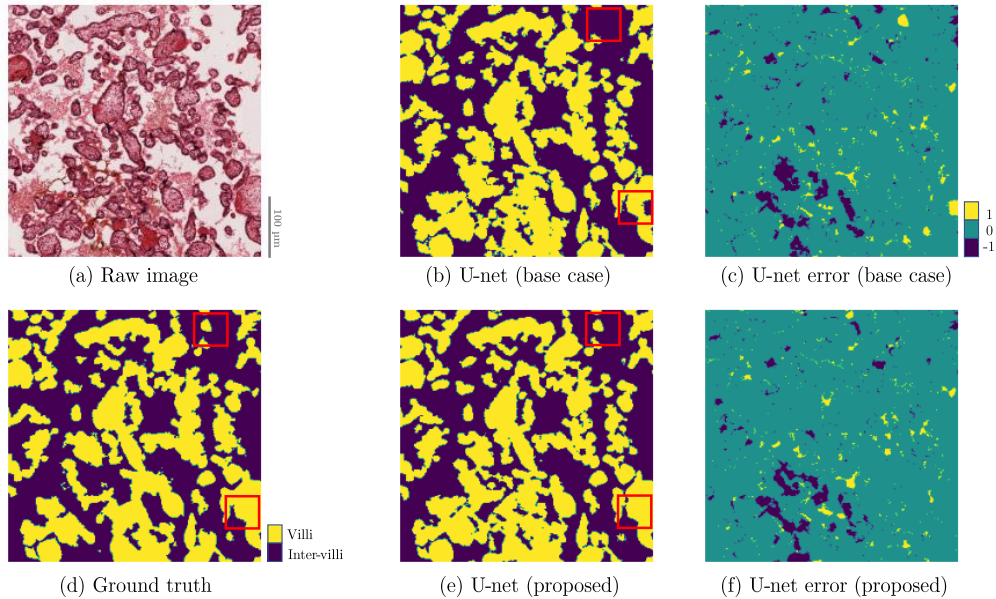} \stepcounter{fig_num}
	\caption{Comparing the predicted masks by training U-net on the augmented datasets with base case approach (b) and the proposed approach (e), as well as the mask prediction errors (c and f) compared to the ground-truth image (d).}
	\label{fig:performance_comp2}
\end{figure}

To look more closely at the segmentation results, we have compared the visual performance of U-net when trained on each of the augmented datasets discussed (Fig. \ref{fig:performance_comp2}). It has been observed that in some cases, such as the one highlighted by a red box on the top right in Figs. \ref{fig:performance_comp2}-d, b, and e, the base case misses the villi islands. Also, in some cases, such as the one highlighted with the red box at the bottom right in Fig. \ref{fig:performance_comp2}-d, the base case approach predicts incorrect holes in the center of the large villus with less color intensity (Fig. \ref{fig:performance_comp2}-b), which does not occur when using the proposed augmentation method (Fig. \ref{fig:performance_comp2}-e). Furthermore, the error maps for each of the cases are presented in Figs. \ref{fig:performance_comp2}-c and f to provide a better understanding. 

\subsection{Morphological comparison}
\begin{figure}[H]
	\centering\includegraphics[page=\value{fig_num},width=.8\linewidth]{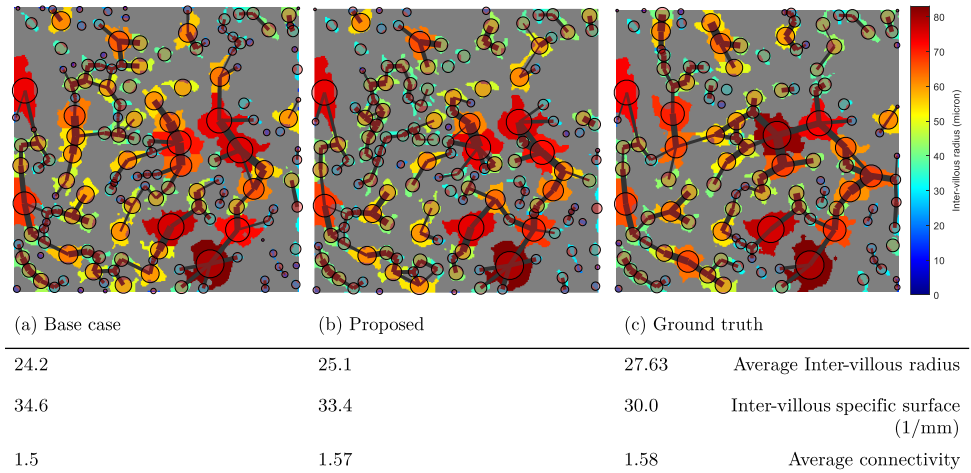} \stepcounter{fig_num}
	\caption{Comparing the morphology of the intervillous space obtained using the base case (a) and the proposed (b) methods of augmentation. Values of average intervillous radius, intervillous specific surface, and average connectivity of the intervillous network of the proposed method are closer to the ground-truth values (c) compared to the base case.}
	\label{fig:morph_compare1}
\end{figure}

To take a further step in comparing the proposed and base-case augmentation methods, we have compared the morphology of the extracted intervillous spaces with the ground-truth data to demonstrate the effects of the proposed augmentation technique on the downstream analyzes. For this purpose, the watershed segmentation algorithm has been used to identify different intervillous chambers and extract the structure of the maternal side internal network. Three morphological features are extracted from the images comprising the average radius of the intervillous chambers, the specific surface of the intervillous space, and the average connectivity of the intervillous network. For all three features studied, it is observed that the proposed method of augmentation led to a closer set of features to the ground-truth (Figs. \ref{fig:morph_compare1} -a, b, and c). On average, the features of the proposed method show around 4\% deviation from the ground-truth values, which is comparable with 7\% relative error when using the base case approach.   

\begin{figure}[H]
	\centering\includegraphics[page=\value{fig_num},width=.8\linewidth]{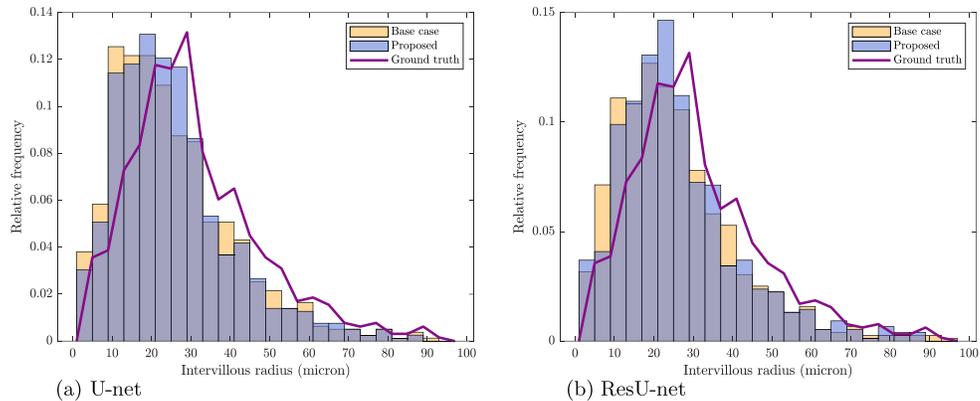} \stepcounter{fig_num}
	\caption{Comparing the intervillous chambers radius distributions obtained from the base case and the proposed methods as well as the ground-truth distributions for HV \#1 using U-net and ResU-net as examples.}
	\label{fig:morph_compare2}
\end{figure}

Furthermore, we have compared the distribution of the intervillous chamber size for the available validation images of HV $\#1$ (Fig. \ref{fig:morph_compare2}). Both the U-net and ResU-net deep learning models give back the chamber size distributions that are more close to the ground-truth distribution when using the proposed augmentation technique. 
 
\subsection{Effect of reconstruction grid size}
\begin{figure}[H]
	\centering\includegraphics[page=\value{fig_num},width=.8\linewidth]{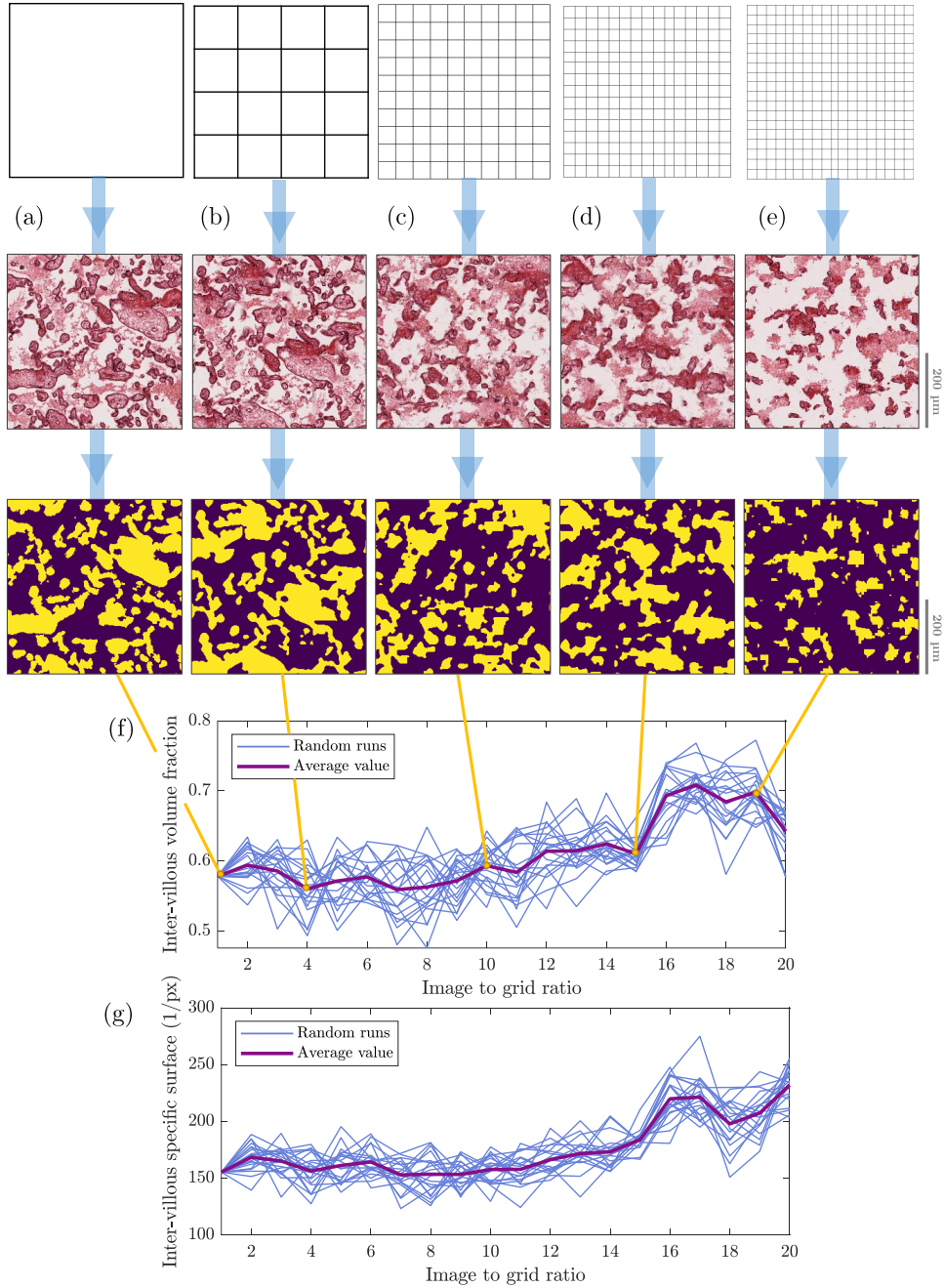} \stepcounter{fig_num}
	\caption{Effect of sampling grid size on image textures and corresponded segmentation masks (a to e), calculated intervillous volume fraction (f) and intervillous specific surface (g).}
	\label{fig:grid_size}
\end{figure}

Size of the reconstruction grid presented in this study is a subject of uncertainty. Therefore, investigating its effect on the diversity of the augmented images could be beneficial. For this purpose, a sensitivity analysis is performed by gradually increasing the image-to-grid-size ratio from 1 to 20 and calculating the morphological features of the images obtained in each step. This process has been repeated 20 times to obtain the average of the existing trends. Fig. \ref{fig:grid_size} shows the effect of the size of the reconstruction grid on the morphological features of the intervillous space. As can be seen, decreasing the size of the grid has led to an increment in the intervillous volume fraction as well as the intervillous specific surface due to the generation of small and separated villi islands, which looks abnormal at the extreme ratio of 20. 
In summary, using a low image-to-grid ratio does not create a significant variation in the augmented dataset, while using a very fine grid could lead to losing the original textures and elements, as was the case in Fig. \ref{fig:grid_size}-e. Consequently, based on the level of detail available in the images, it is recommended to select a moderate grid size ratio after visual inspection, which in the present study was equal to 6.

\subsection{Future applications}

\begin{figure}[H]
	\centering\includegraphics[page=\value{fig_num},width=.8\linewidth]{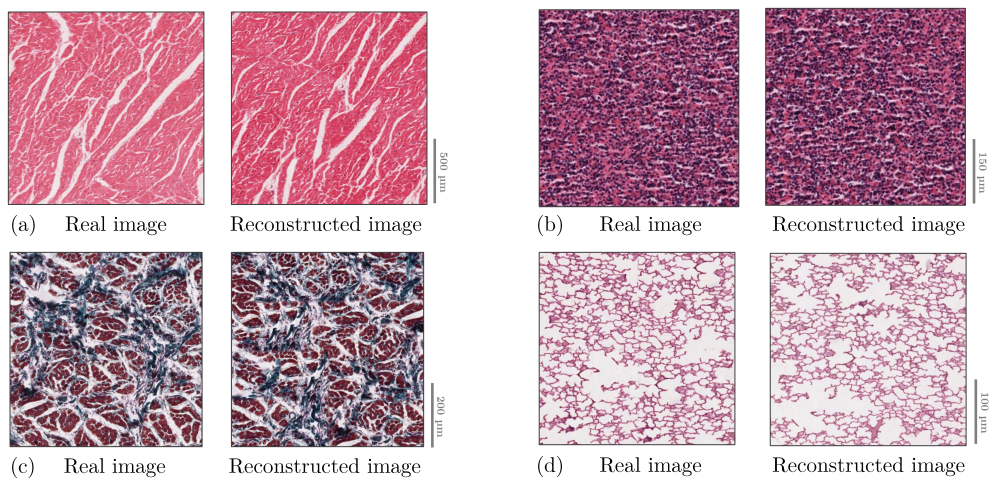} \stepcounter{fig_num}
	\caption{Four examples of using the proposed method of image reconstruction for making realizations of different histological images, a) H\&E image of human myocardium tissue from heart left ventricle, b) H\&E image of human cerebellum granular layer c) H\&E image of human palate spongy bone from sagittal section, d) H\&E image of human Alveolar cells, and lumens of pulmonary capillaries. Reconstructed images do not only mimic the local textures but also imitate global features like the oblique presence of the white connective tissues and capillaries in the myocardium example (a).}
	\label{fig:future_applications}
\end{figure}

The presented augmentation method has successfully generated synthetic realizations of the original histological images that are indistinguishable from the real images when looking with the naked eye. This approach can be useful in the reconstruction of a wide range of histological images, as presented in Fig. \ref{fig:future_applications}. Fig. \ref{fig:future_applications}-a illustrates a real and normal H\&E image of human myocardium tissue from the left ventricle of the heart, as well as a reconstructed example using the grid ratio of 6. A remarkable observation is that the proposed method is capable of maintaining directional textures, which are sometimes clinically significant. Furthermore, in another example, Fig. \ref{fig:future_applications}-b shows a pair of real and reconstructed H\&E images of the human cerebellum granular layer with a horizontal texture that has been maintained after reconstruction. The reason for this capability is that the proposed reconstruction technique does not use flipped sub--sections unless it is necessary due to the scarcity of similar non-flipped sub-sections (refer to Fig. \ref{fig:workflow}-c). Finally, two other examples provided show a histological image of human palate spongy bone from the sagittal section and human alveolar cells, respectively (Fig. \ref{fig:future_applications}-c and d). In these examples, the resemblance of the reconstructed images with the real histological images is noticeable, and it is recommended to run a blind test for experienced pathologists to distinguish between real and synthetic images in future studies. 

\section{Conclusions}
\label{sec:con}
In this study, a new method of histological data augmentation has been presented, which is suitable for the condition with minimal availability of labeled data. We have compared the proposed method of augmentation with existing common approaches in the literature and found it to be effective for training deep learning segmentation models on histological images of the human placenta. However, applications of the proposed method may not be limited to placental histological images, and it is recommended that other types of tissues be investigated in future studies. In summary, the main conclusions and findings of this study are as follows. 
\begin{itemize}
	\item  The proposed data augmentation technique creates a wider range of reconstructed images with higher degree of diversity using a single training image compared to the base case augmentation technique, which is common in the literature. Based on the dimension--reduced maps of the two datasets, we have found the domain covered by the proposed method to be 5 times larger than the base case method of augmentation, which in our tested example gave us a higher chance of successful prediction for out-of-sample images.  
	
	\item  The proposed method led to an improvement in the performance of the four deep learning models tested. The binary cross-entropy validation loss has decreased by 42\% when using the porposed method. In addition, the presented approach helped to keep the learning curves more stable and less noisy. In other words, using the presented approach, a deep learning model reaches a more reproducible state after passing a certain number of training epochs. 
	
	\item  The proposed reconstruction technique is capable of maintaining directional image textures while generating new synthetic realizations that are visually difficult to distinguish from real images.
	
\end{itemize}
\section*{Availability of the code and data}
Please refer to \url{www.github.com/ArashRabbani/Augmented-Pattern} to access the code and instructions for reconstruct new realizations.

\bibliography{ref}

\end{document}